# HERITAGE: A MONTE CARLO CODE TO EVALUATE THE VIABILITY OF INTERSTELLAR TRAVELS USING A MULTI-GENERATIONAL CREW

**FRÉDÉRIC MARIN**
*Université de Strasbourg, CNRS, Observatoire astronomique de Strasbourg, UMR 7550, F-67000 Strasbourg, France*
Email: frederic.marin@astro.unistra.fr

To evaluate the feasibility of long duration, manned spaceflights, it is of critical importance to consider the selection and survival of multi-generational crews in a confined space. Negative effects, such as infertility, overpopulation and inbreeding, can easily cause the crew to either be wiped out or genetically unhealthy, if the population is not under a strict birth control. In this paper, we present a Monte Carlo code named HERITAGE that simulates the evolution of a kin-based crew. This computer model, the first of its kind, accounts for a large number of free parameters such as life expectancy, age range allowed for procreation, percentage of infertility, unpredictable accidents, etc... to be investigated proactively in order to ensure a viable mission. In this first paper, we show the reliability of HERITAGE by examining three types of population based on previously published computations. The first is a generic model where no birth/population control has been set up, quickly leading to fatal overcrowding. The second is the model presented by Moore (2003), that succeeds to bring settlers to another Earth under a 200 year-long flight, but the final crew is largely diminished (about a third of the initial crew) and about 20% of them show inbreeding of various levels. The third scenario is the model by Smith (2014) that is more successful in maintaining genetic diversity for the same journey duration. We find that both the Moore and Smith scenario would greatly benefit from coupling a kin-based crew together with a cryogenic bank of sperm/eggs/embryos to ensure a genetically healthy first generation of settlers. Finally, we also demonstrate that if initial social engineering constraints are indeed needed to maintain an healthy crew alive for centuries-long journeys, it is necessary to reevaluate those principles after each generation to compensate for unbalanced births and deaths, weighted by the inbreeding coefficient and a need for maximizing genetic diversity.

## 1. INTRODUCTION

With the increasing number detection of exoplanets (3660 planets as of August 24th, 2017), the possibility to find an Earth-like telluric world is expanding. The recent detection of a close[1] terrestrial planet around the red dwarf Proxima Centauri is of utmost interest as the planet's equilibrium temperature is within the range where water could be liquid on its surface [1, 2]. Situated at a distance of 1.295 parsecs [3], Proxima Centauri b is thus a perfect target for a short interstellar travel. At one percent of the speed of light it requires a 423 year-long journey. But, even considering a speed that is well in excess of what we can currently achieve, it would still require more than one generation to reach Proxima Centauri b.

One of the immediate consequences of this distance is that travels to Proxima Centauri b cannot be achieved within the life expectancy of a human. It requires a long-duration space voyage, which necessitates to find a solution for the crew to survive hundreds of years in deep space. Despite advances in the field [4], cryogenic technologies are not yet viable solutions as freezing cells will create ice crystals which break down cell walls as they expand (vitrification), reducing the body to mush once it is warmed up again. Suspended animation scenarios, where the physiological functions of the crew members are slowed down until arrival, have yet to be invented. To overcome the problem of the crew's survival, embryo ships were investigated [5].

Building genetic banks of frozen early-stage human embryos may be an alternative option to slowly populate an Earth-like exoplanet, where robotic equipment would help the embryos to mature and develop once the destination is reached. By doing so, the size of the ship and its speed can be reduced to feasible projects. For many years the lack of working artificial wombs was hampering this possibility but recent development in extra-uterine systems to physiologically support extreme premature animals (here a lamb) are promising [6].

The best option might be to rely on giant self-contained generation ships that would travel through space while their population is active. This is the most well-understood option: population genetic diversity over time can be modelled, future developments in technology cannot. Several ideas for structures and designs are presented in [7, 8, 9] but they lack a mathematical and statistical examination of their hypotheses. The anthropologist John Moore was the first to use an ethnographic tool named ETHNOPOP to rigorously calculate the minimum number devising a multi-generational journey [10]. ETHNOPOP is a stochastic, open-source code[2] that simulates the marital and demographic situation facing small colonizing bands of human being [11]. It uses external modules to episodically create epidemics and disasters, but those modules were never used in the context of a spaceflight. This software was made to compute and analyze the historical migrations of early human groups by taking into account their demographic trends. ETHNOPOP incorporates four demographic variables linked to death and birth risks as well as three cultural variables (marriage choice, polygyny, marriage pool). Considering a space journey where immigration and emigration are not possible, Moore concluded that a 200 year-long mission should have an initial crew of 150 – 180 people. According to him, the crew should be young and allowed to procreate only late during the women's reproduction life in order to postpone the appearance of the first generation for as long as possible. This is a safety condition to prevent overpopulation. Theses numbers were also selected to avoid too much inbreeding. More recent calculations, including the effects of mutagenesis, genetic migration, mate selection and birth drifts, are revising these numbers upwards. Cameron Smith published an estimation for a genetically viable population to endure a 5-generation space voyage under strict population genetic control [12]. Among other results, it was found that an initial crew of 14000 – 44000 members is well-optimized to ensure healthy offspring, even in the case of a sudden disaster occurring once during the mission. According to his study, a crew of 150 people would always be on the verge of extinction in the case of a large-scale catastrophe. Smith advocates for a much larger gene pool, which translates to larger crews. The important variation in the estimate of the minimum size is due to the underlying hypotheses used by the author, who calculated the final numbers using a simple statistical approach. Thus, it appears that estimating a secure number for the initial population is challenging, even if we do not account for the psychological effects that the loss of the home planet can have on the initial crew [13].

---

1. Proxima Centauri, also known as α Centauri, GL 551 or HIP 70890, is the closest star to the Sun. Any habitable world around Proxima Centauri is thus the closest exoplanet system humanity can find.

2. The source code is unfortunately no longer available.

In this first paper, we introduce a new Monte Carlo code that aims to constrain the input parameters (i.e. the selection of the initial crew) and define the social engineering rules needed for a centuries-long space journey. This simulation tool, the very first Monte Carlo method solely dedicated to kin-based spatial simulations, can be entirely fine tuned by the user using a larger number of input parameters than any of the codes previously presented in the literature. HERITAGE calculates the evolution of crew generations, accounting for life expectancy, inbreeding and multiple random parameters that will be listed in the following section of this paper, to check whether the final population landing on a terrestrial, humanfriendly planet is genetically healthy. Section 2 will present the numerical tool and the required input parameters. Section 3 will investigate first a random, uncontrolled population (Sect. 3.1), then explore the numbers suggested by Moore (Sect. 3.2). In Sect. 3.3, we will proceed with the recent results from Smith. Conclusions and future work are presented in Sect. 4.

## 2. THE HERITAGE CODE – AN OVERVIEW

HERITAGE[3] is a Monte Carlo code written in C/C++. The Monte Carlo Method relies on repeated random sampling to solve problems that might be deterministic in principle [14]; this technique is particularly well suited for this type of simulation as the random processes involved in any event occurring in the spaceship can be put into numbers. The method allows investigation of the possible outcomes of a series of unpredictable situations in order to assess the impact, allowing for better decision making under uncertainty [15]. Probabilistic techniques are commonly used in astrophysical simulations where millions of random processes (such as light scattering) govern the observed properties of stars, nebulae or galaxies [16, 17]. A Monte Carlo code allows the individual simulation of each crew member rather than applying statistics on a large group. Since HERITAGE is written in C++, the "Human" class[4] creates a different object (i.e. a member of the crew) every time the class constructor is called. In doing so, each crew member is unique as all of his/her characteristics are randomly determined according to the input constraints selected by the user. Here we apply the Monte Carlo routine to the random processes that will impact the life of the multi-generational crew members, such as life expectancy, number of children, age of death, risk of deadly accident and other parameters that are listed in Table 1.

### 2.1 Parameters for the Ship

Ideally, the duration of the interstellar travel is to be set according to the velocity of the space ship and the distance from Earth to the new planet, accounting for a deceleration phase when the ship is in the stellar system neighbourhood. The Solar Probe Plus [18], a NASA mission scheduled to launch in 2018, will be the fastest human-made object to wander space at orbital velocities as high as 724000 km.h$^{-1}$ (i.e. 0.002$c$). For perspective, at this speed, it would take 6300 years to reach Proxima Centauri b. We will leave an exact estimation of the duration for another study and now arbitrarily fix the duration of the flight to 200 years in order to match the duration of the different scenarios for kin-based crew described in literature [10, 12].

The colony ship capacity is function of the design of the vessel, its storage capacity and facilities (gardens, social areas, internal manufacturing ...). In this topic different projects, such as *Hyperion* (parent organization Icarus Interstellar [19, 20]), and the *100 Years Starship* project [21, 22, 23], seem to be the most advanced in the development of a concept for manned interstellar missions. So far, the code only gives a warning about the onset of overpopulation but let the simulation ends. It is at the discretion of the user to turn off the code when overcrowding happens.

It remains an open question as to whether intelligence is determined primarily by nature or nurture, and if the ship operations are complex (as in present day spacecraft) then it is by no means clear if the necessary skills can be guaranteed to be taught to each successive generation. For this reason, for the rest of this paper we neglect such issues as ship operation and maintenance. This effectively assumes the ship's operation is entirely automatic or requires only very simple human intervention.

### 2.2 Parameters for the Initial Crew

The initial crew is composed of a user-defined ratio of women and men, ideally 50% – 50% to maximize genetic diversity. The age of the crew members should be relatively young in order to postpone as much as possible the first generation of space-born humans [10, 12]. The age range is computed using a normal (Gaussian) function. The probability density of the normal distribution is:

---

3. "Heritage" means "Legacy" in French, in the sense that each generation of crew members lives only to bequeath their knowledge and a common goal to the new generation: reaching the target exoplanet.

4. A class in C++ is a user defined type or data structure declared with keyword class that has data and functions. Defining a class is similar to defining a blueprint for a data type. This doesn't actually define any data, but it does define what the class name means, that is, what an object of the class will consist of and what operations can be performed on such an object.

**TABLE 1:** *Input Parameters of the Simulation.*

| Parameter | Value | Units |
|---|---|---|
| Number of space voyages to simulate | 100 | (integer) |
| Duration of the interstellar travel | 200 | (years) |
| Colony ship capacity | 500 | (humans) |
| Number of initial women | 75 | (humans) |
| Number of initial men | 75 | (humans) |
| Age of initial women | 20/1 | (years) |
| Age of initial men | 20/1 | (years) |
| Women infertility | 0.10 | (fraction) |
| Men infertility | 0.15 | (fraction) |
| Number of child per women | 2/0.5 | (humans) |
| Twinning rate | 0.015 | (fraction) |
| Life expectancy of women | 85/15 | (years) |
| Life expectancy of men | 79/15 | (years) |
| Mean age of menopause | 45 | (years) |
| Start of permitted procreation | 35 | (years) |
| End of permitted procreation | 40 | (years) |
| Chances of pregnancy after intercourse | 0.75 | (fraction per year) |
| Initial consanguinity | 0 | (fraction) |
| Allowed consanguinity | 1 | (fraction) |
| Life reduction due to consanguinity | 0.5 | (fraction) |
| Chaotic element of nay human expedition | 0.001 | (fraction) |
| Possibility of a catastrophic event | 1 | (boolean) |
| Year at which the disaster will happen | 75 | (year; 0 = random) |
| Fraction of the crew affected by the catastrophe | 0.30 | (fraction) |

The $\mu/\sigma$ values shown for certain parameters indicates that the code needs a mean $(\mu)$ and a standard deviation value $(\sigma)$ to sample a number from a normal (Gaussian) distribution.

$$f\left(x|\mu,\sigma^2\right) = \frac{1}{\sqrt{2\pi\sigma^2}} e^{\frac{-(x-\mu)^2}{2\sigma^2}}$$

where $\mu$ is the mean of the distribution (and also its median and mode), $\sigma$ is the standard deviation and $\sigma^2$ is the variance. Age is sampled using a random number generator. The same process applies for the life expectancy of the crew members, with global statistics showing a longer life expectancy for women [24]. The standard deviation in life span was estimated after age 10 and found to be about 15 years in the US and France (13 years in Sweden and Japan), see [25].

### 2.3 Parameters for Procreation

Each human is characterized by a fertility fraction that represents their natural capability to produce offspring. This value may be different for males and females. The fertility or lack of fertility is randomly set when creating the crew member in the software. The number of children a fertile woman can have during her lifetime is chosen according to another normal (Gaussian) distribution. HERITAGE accounts for at least one year between two pregnancies and may randomly create twins instead of a single baby. At this stage, the twin offspring are dizygotic (they develop from two different eggs, as opposed to identical twins coming from the same egg, for whom the frequency of occurrence is much lower) and the percentage of twins in the population is set to 1.5% according to recent demographic reviews [26]. Women stop being productive at a given menopausal age, to be determined by the user (here set to the world average value of 45 years, see [27]).

If we want to control the population level to ensure genetic diversity and avoid overcrowding, it is possible to state at which age the procreation period should start and finish[5]. Those limits apply to both men and women. If no thresholds are defined, the default value sets the start of the procreation age at 18 (threshold of adulthood in the largest number of countries) and the end at the menopause age. It is known that pregnancy is problematic at later ages but, for simplification, we did not account for this dilemma. Note that it is not necessarily a problematic question if one considers in vitro fertilization, see Sect. 4. Finally, each breeding event is not guaranteed to create a new human being and the percentage of successful births can be also controlled (and related to our previous point about late births). A fertile couple has about 10% chances of a pregnancy if they mate once per month. For more regular intercourses this threshold is higher, increasing to 30% when intercourses occur twice a week [28]. In the case of an intensive reproduction program, the chances of pregnancy after intercourses gets higher and was arbitrarily fixed to 75% (user-adjustable value).

---

5. To do so, contraception is the most effective method.

Consanguinity due to inbreeding is a serious problem in closed systems where long-term population planning is an important key to success [29]. The code is not wired to account for moral questions about inbreeding and, if the user does not put a threshold on the maximum inbreeding permitted aboard, can let any male and female procreate randomly, leading to potential consanguinity as the couples can be polygamous. However, it is not genetically advisable to cross sibling-father-daughter, or mother-son, i.e. any combination that leads to more than 20% consanguinity. Such a rate of inbreeding can result in a weak gene pool and immune weakness, leading to the abnormal development of children [30, 31]. The coefficient of consanguinity $F$ is the probability that the two allele genes that an individual has at a locus are identical by descendance, and it is calculated using the equation presented by Wright [32]. $F$ is equal to the sum of inbreeding induced by the common ancestors:

$$F = \sum_{i=1}^{n} C_i$$

where $n$ is the number of common ancestors. For the remainder of this paper, $F$ will stand for "consanguinity coefficient" or "inbreeding coefficient", both terms being interchangeable. The coefficient $C_i$ for a given common ancestor $i$ is calculated as follows:

$$C_i = \frac{1}{2^{P+M+1}}$$

where $P$ is the number of generations between the father and the common ancestor, and $M$ is the number of generations between the mother and the common ancestor. We see that the greater the remoteness, the more limited the risks are, as each generation $i$ has a probability 1/2 of transmitting this allele to $i + 1$; therefore a probability $(1/2)^n$ after $n$ generations. The coefficient of consanguinity is saved for each crew member, so in the case a couple has a common ancestor, the coefficient $(1 + F)$ is assigned to the calculation of the coefficient of inbreeding to account for the initial consanguinity of this peculiar ancestor. The deleterious effects of inbreeding begin to become evident at $F > 5\%$. At $F > 10\%$, there is significant loss of vitality in the offspring as well as an increase in the expression of deleterious recessive mutations that eventually drives a population to extinction [33, 34]. For this reason, the user can fix a maximum consanguinity threshold within the crew members. This then prevents the code from selecting partners of similar bloodlines, avoiding incest between family members.

Properly calculating the coefficient of consanguinity for each crew member allows HERITAGE to avoid using the rule of thumb invented by Franklin and Soule [35, 36] and called the "50/500" rule. The basics of this rule is that, to survive for 100 years in a closed environment (such as a starship or a non-populated planet), a minimum of 500 randomly chosen settlers or 50 hand-picked settlers all who are unrelated and of breeding age, are mandatory. For longer durations, the rule has a linear scaling. We opted for a proper method that will validate the "50/500" rule in the next section.

### 2.4 Parameters for the Journey

Few major catastrophic events can happen during the flight. Interstellar comets (comets located in interstellar space that are not gravitationally bound to a star [37]) and interstellar asteroids are not frequent enough (interstellar number density 1.4 $10^{-4}$.AU$^{-3}$, with AU = 1.496 $10^8$ km) to represent a real risk of collision [38], taking into account that any interstellar ship would necessarily be equipped with collision detectors. Aside from technological and mechanical failures, the most harmful situation can only occur within the crew, where a major disease could break out. Smith [12] discussed the possibility of a sudden disaster, sweeping about 30% of a restricted population such as in the years 1346 – 1353, when medieval Europe was facing the Black Plague [39]. Following his recommendation, we included the possibility to have a catastrophe which can destroy a user-specified percentage of the crew. The emergence of the disaster is chosen randomly by the program if the input parameter is set to zero, otherwise the date is decided by the user. Additionally, a chaotic element, representative of any human expedition (deadly accident, premature death, serious illness ...) was also implemented and is at the discretion of the user. The chaotic element is different from the sudden disaster: whereas the catastrophe kills a large fraction of the population (but only once) the chaotic element is responsible for premature deaths of individuals throughout the journey.

### 2.5 A Year in the Vessel Step-by-Step

At the beginning of the program, HERITAGE reads the input file where all the parameters for the mission are listed (input. dat). Then, the code individually creates the crew members by calling the Human C class. It gives an identification number to each member, selects their sex according to the specified ratio, randomly assigns their age, selects their date of death following the Gaussian life expectancy, checks if the crew member is fertile or not (boolean value selected according to a random drawing compared to the percentage of infertile population) and sets their initial consanguinity (0 if the crew was carefully selected). If the crew member is a woman, the code also computes the number of children the member can have.

After this initializing phase, HERITAGE loops the same series of instructions each year after launch. First it checks if a catastrophic, sudden disaster happens and acts accordingly by reducing (by a certain percentage) the number of crew members. Then, the code checks whether each crew member has reached their age of death. If so, it withdraws the crew member but keeps in memory all of its characteristics for further statistical analyses. Then the chaotic seed is applied and can randomly kill a crew member for unpredictable reasons (cerebrovascular accident, accidental death ...). The next step is the procreation loop that will randomly decide (according to the input parametrization) if a woman will give birth during the year. HERITAGE finds a breeding partner, evaluates the possibility for the couple to have a child then calculates the consanguinity of the offspring. Procreation will happen if the consanguinity is lower than the input threshold. The code randomly chooses if there will be twins and also chooses the sex of the child (50% chance to be a girl). Then HERITAGE creates a new object from the class Human and assigns all the necessary information to this new crew member. If the consanguinity of the offspring is larger than 10%, the life expectancy of the child is reduced due to the inbreeding depression phenomena by a factor determined by the user (here set to 50%, but the real fraction is unclear for humans, see [40] for inbreeding depression in a butterfly metapopulation).

When all the crew members have been evaluated in this loop, the age of all crew members is increased by one and another year of travel starts until the completion of the main loop. A flowchart of the code is presented in Fig. 1.

HERITAGE saves all the characteristics of the living and dead members of the ship then automatically computes several key numbers that are plotted in graphs. This includes the crew evolution in terms of age and sex, the number of birth/death per year, the distribution of the crew age per year plotted in a 3D density plot (plus a 2D slice of this plot at different epochs) and the level of consanguinity of the crew (maximum, minimum, average and variance). The code ends by stating the running time of the simulation, typically less than a minute for a single 200 year-long trip with 150 crew members. One can loop the results over dozens, hundreds or thousands of trips to obtain better statistics.

## 3. SIMULATIONS

In this section, we present the HERITAGE results for three different cases. The first one concerns an interstellar mission without any control on the population level. The second simulation tests the numbers suggested by Moore [10] and the third simulation investigates the predictions from Smith [12]. For all simulations, the duration of the flight was set to 200 years and includes an equal number of men and women. The age of the initial crew members follows a normal distribution centered around 20 years old. The simulations were looped over 100 voyages to obtain sufficient statistics. Each section will describe the changes in the resulting parameters. The parametrization for each scenario is summarized in Table 2.

### 3.1 Results for a Uncontrolled Population

It is an ethical and moral question as to whether a permanent control over the breeding selection and the number of child can be accepted by a multi-generational crew. In the case of an absence of control, the number of children per women might be more important and we fixed the default parameter to a mean of 3 children (with a standard deviation value of unity). By using a Gaussian distribution of children per woman, we ensure that there is a reasonable number of offspring per woman, avoiding illogical situations where the crew would constantly seek for reproduction. However, we let the size of the population and its genetic diversity uncontrolled, so that there is no enforced regulation of breeding and no self-imposed lower limits by the crew either. We allowed the crew members to breed as soon as they turn 18 years old until the natural limit of menopause. We stress that this is something of a worst-case scenario: the crew aren't concerned with managing resources and have no taboos against incest, which is not realistic. This is a purely theoretical scenario to see how fast the mission could fail without strict controls. See the parametrization in Table 2, first column.

Figure 2 (left) shows the evolution of the population within the interstellar ship over 200 years. It appears that the number of crew members steadily increases over time until a sudden disaster reduces the populace by a third. This is due to the sudden disaster that happened at the 75th year of travel. The population soon recovers as nothing prevents breeding so the number of offspring continues to rise. The growth of population is almost exponential at the end as nothing, except inbreeding (not restricted in this section), can stop the multiplication of humans. It is interesting to note that our results are in perfect agreement with the biological exponential growth observed in natural habitats where a population of an organism living in

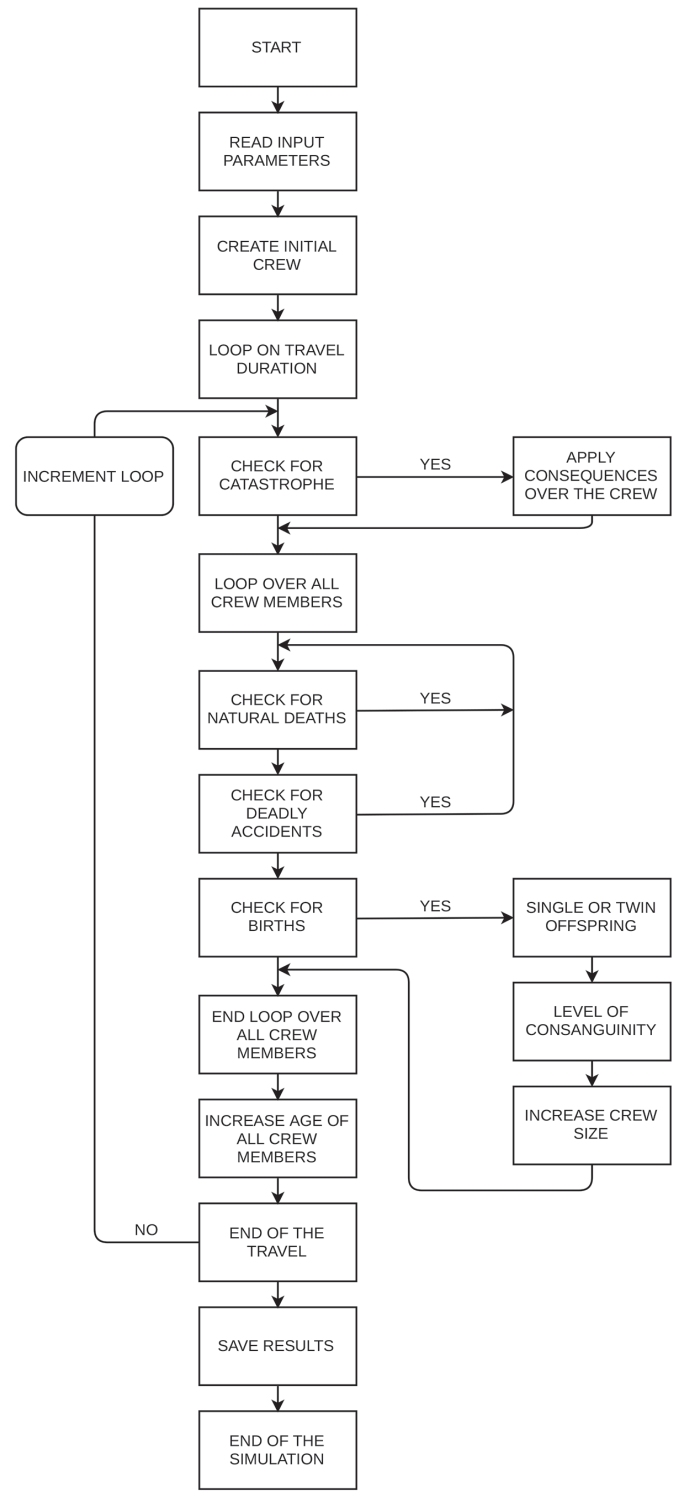

**Fig. 1 Flowchart of the program. Each box meets the conditions from the input parametrization defined by the user. HERITAGE runs until the end of the travel period, regardless whether the crew is still alive.**

the habitat grows in an exponential or geometric fashion if the resources availability is unlimited. A mathematical description of biological exponential growth is presented in Miller & Harley [41] and our results are clearly within this exponential regime. The colony ship capacity, shown in dotted line, is soon exceeded despite the disaster. The sudden disaster clearly appears on Fig. 2 (right), where the number of births and deaths follow the trend of the population growth at different rates. There are almost no deaths (a few per years due to the chaos parameter) until year 50 of the journey, when old age naturally

**TABLE 2:** *Input Parameters of the Three Simulations.*

| Parameter | "Uncontrolled" | "Moore" | "Smith" |
|---|---|---|---|
| Nb. travels | 100 | - | - |
| Duration | 200 | - | - |
| Capacity | 500 | 500 | 50000 |
| Nb. women | 75 | 75 | 7000 |
| Nb. men | 75 | 75 | 7000 |
| Age women | 20/1 | - | - |
| Age men | 20/1 | - | - |
| Women inf. | 0.10 | - | - |
| Men inf. | 0.15 | - | - |
| Nb. child | 3/1 | 2/0.5 | 2/0.5 |
| Twinning | 0.015 | - | - |
| Life exp. women | 85/15 | - | - |
| Life exp. men | 79/15 | - | - |
| Menopause | 45 | - | - |
| Start proc. | 18 | 35 | 35 |
| End proc. | 45 | 40 | 40 |
| Chances preg. | 0.75 | - | - |
| Initial cons. | 0 | - | - |
| Allowed cons. | 1 | - | - |
| Life red. cons. | 0.5 | - | - |
| Chaos | 0.0001 | - | - |
| Catastrophe | 1 | - | - |
| Year disaster | 75 | - | - |
| Fraction affected | 0.30 | - | - |

"-" indicates that the same numbers are used for the three simulations.

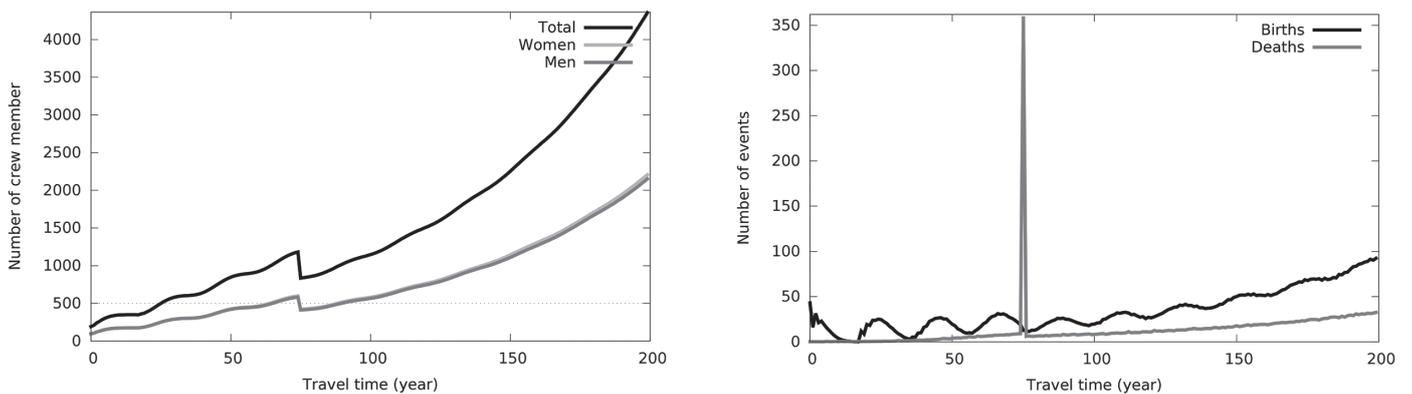

**Fig. 2** (Left) Crew evolution in terms of population number for a 200 years long trip where no birth control is applied. (Right) Number of births and deaths per year. The sharp peak is due to a catastrophic sudden disaster. The dotted line is the colony ship capacity.

reduces the population. The birth rate increases with time but not steadily as the number of couples is relatively small at the beginning (a maximum of 75 couples for the first 18 years). This naturally creates a Gaussian distribution of births in the first years of travel that repeats over time until the end of the mission. In comparison, the number of deaths follows an almost linear increase over time, never equating the birth rate.

The age clustering can be seen in Fig. 3 for men (top), women (middle) and the whole group (bottom). The x-axis represents the age of the population, the y-axis is the duration of the interstellar trip and the number of crew members of a given age at a given time is colour-coded. White colour indicates zero members of this age. The age clustering is very sharp at the beginning of the journey as the initial crew had a distinct age group. As soon as their children reach the minimum age for procreation, the distinct demographic echelons become less clear and the separation between age groups vanishes after the second generation. On the 2D slices of the clustering presented in Fig. 4, the destruction of the demographic echelons appears clearly between the onset of the simulation and the keystone time of half the mission. There is still a remnant of the age clustering but the generations on-board the ship cover the whole range of possible ages. The maximum age of the oldest crew member was just under 130 years in this simulation. At the end of the mission, no structure can be seen in the plot and the birth rate is about two times larger than the death rate, explaining the exponential demographic explosion seen in Fig. 2 (left).

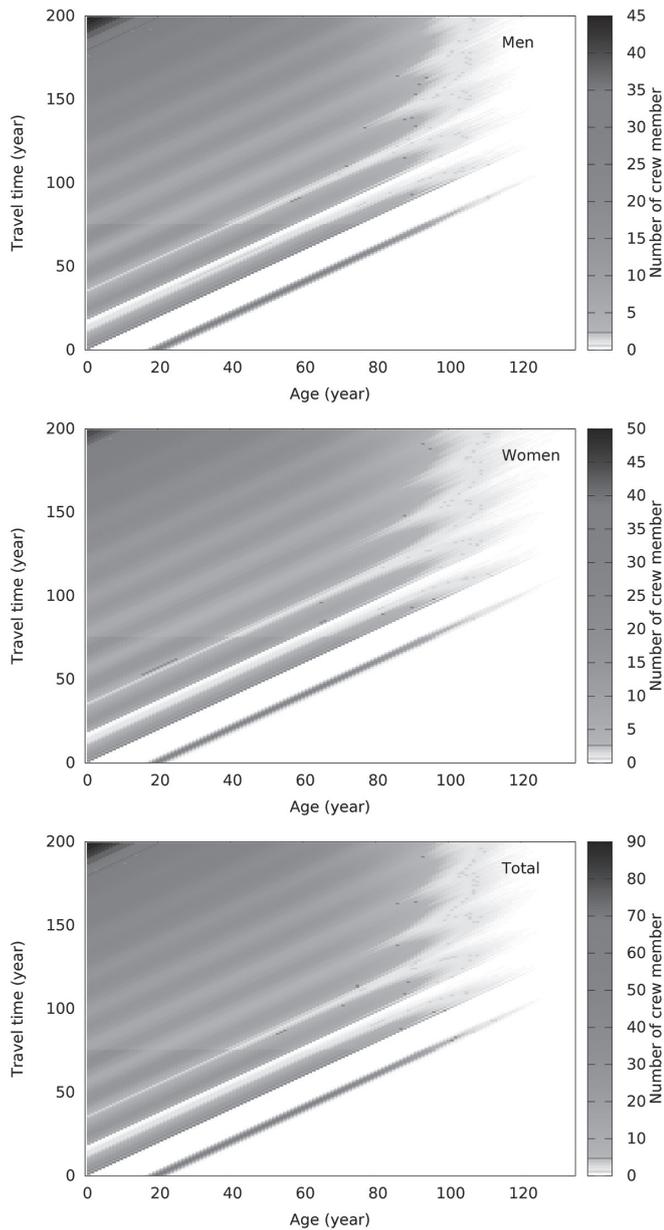

**Fig. 3** Density of crew members of a given age (x-axis) over time (y-axis) for a 200 years long trip where no birth control is applied. The population density is colour-coded from 0 (white) to maximum (black). Top: Men; Middle: Women; Bottom: Total.

Figure 5 shows that the limited number of crew members at the beginning of the mission drives the emergence of consanguinity within the offspring. The first generation of crew members born inside the vessel starts to reproduce regardless of the family affiliation and the inbreeding coefficient reaches a high value (21.00%) as soon as 30 years after launch. The sudden disaster had no impact on the consanguinity level, which only decreases when the exponential growth of the population begins. The pool for breeding being larger every year, the chances to crossbreed decreases (but due to hereditary consanguinity remains as high as 12.52% until the end of the mission). However, the average consanguinity within the crew (measured from only those who show a non-zero coefficient) is lower, < 6%, indicating that there is only a very few number of highly consanguineous offspring. The fraction of crew members showing inbreeding confirms this finding as it reaches a maximum of only 2.7% of the total crew (see Fig. 5).

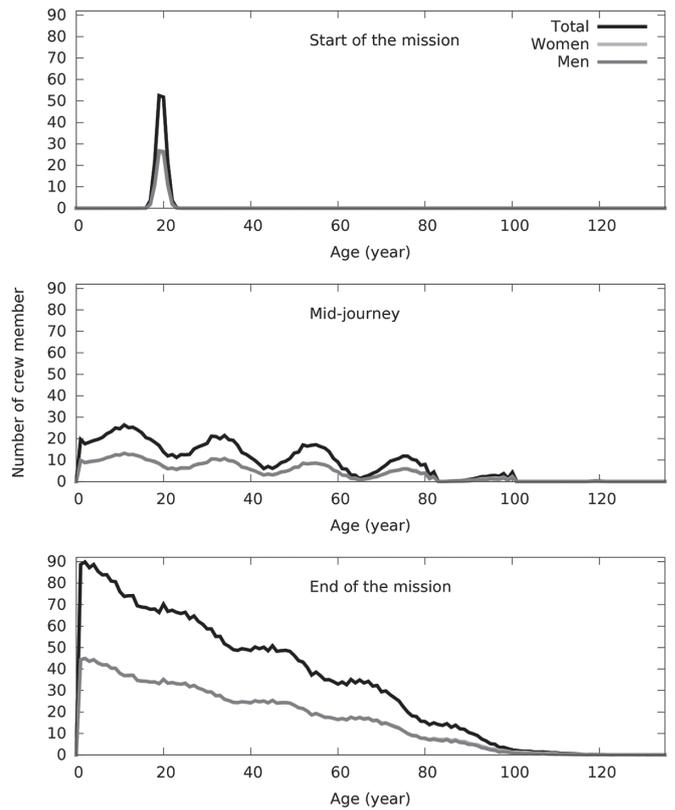

**Fig. 4** 2D slices of the density diagrams shown in Fig. 3 at three different epochs. (Top) The onset of the mission; (Middle) Half of the journey; (Bottom) At the end. Model: 200 years long trip where no birth control applied.

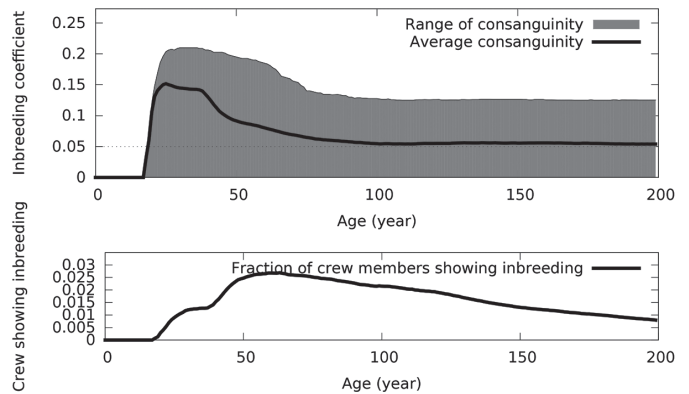

**Fig. 5** Inbreeding within the crew for a 200 years long trip where no birth control is applied. (Top) Inbreeding coefficient as a function of time. The range of consanguinity (maximum-minimum) is shown in grey and the average consanguinity factor $F$ per crew member is shown using the solid black line. The mean is measured from only those who show a non-zero co-efficient. The dotted line represents the limit where deleterious effects onset. (Bottom) Fraction of the crew members showing a non-zero consanguinity.

In conclusions, an uncontrolled crew is likely to develop at a dangerous rate, inducing potentially high consanguinity. Additionally, there is a dissolution of the demographic echelons, which is not as problematic here as it will be in the scenarios to be tested in the next paragraphs. Because of the lack of self-imposed limits by the crew, the colony ship can be overcrowded within 25 years, driving the mission to a fatal end due to the probable development of diseases, a scarcity of food and internal conflicts due to overpopulation and loss of personal space. We agree with the original findings from Moore: the mission necessitates a control of birth rates in order not to saturate the space ship. Postponing parenthood until late in the women's reproductive life is one viable option to delay overcrowding.

### 3.2 Results for a Moore-Like Population

Based on ETHNOPOP modeling, Moore speculated that a population of 150 – 180 people could survive a 200 years long travel if their reproductive cycle was restricted to a finite number of child per woman. In this case, the couples were only allowed to reproduce at an advanced age, clustering the population into discrete age groups, limiting the number of non-productive people within the vessel. Additionally, age clustering helps to maintain genetic variation by lengthening the generations, resulting in smaller sibships. We included the numbers suggested by Moore (i.e. a crew of 150 people with about 2 child per women, see the parametrization in Table 2, second column) and ran HERITAGE.

We find that the social engineering principles of Moore work well: births are clearly identified as peaks in the demographic spectra, compensated by a certain number of deaths shortly after. However the birth/death equilibrium is not maintained using those strict numbers. The number of crew members decreases slowly with time, only slightly impacted by the catastrophic event predicted by Smith (see Fig. 6). Contrary to what Smith postulated [12], a Moore-like population can absorb the loss of 30% of its human population on a short period of time. However, if the social engineering principles are not adapted, the crew is doomed to extinction for longer trips. Nevertheless, at the end of the journey, there is still a crew of about 56 people, with almost the same number of males and females (29 women and 27 men on average).

Clustering in terms of ages is very clear in the 3D (Fig. 7) and 2D (Fig. 8) plots. Gaussian profiles persist from generation to generation with a decreasing amplitude such as seen from previous plots. However the demographic echelons are continually well-defined and effective. The sudden disaster truncates the total amount of crew members in the 3D plot but does not modify the age system. From this point-of-view, a Moore-like parametrization is very effective to maintain an healthy age balance.

The biggest problem concerns the appearance of inbreeding at high levels during the interstellar travel. As seen in Fig. 9, age clustering helps reducing the average level of consanguinity: since procreation can only happen when a settler is at least 35 years old, and lasts for 5 years, it becomes impossible for a daughter to mate with her father, as the father will be well beyond the allowed procreation window. However, inbreeding due to cousin/cousin or brother/sister breeding is still possible, together with other combinations. Hence, the maximum inbreeding coefficient reaches 16.51% at the emergence of the second generation of space-born children. The coefficient remains high during the last years of the journey, with an average consanguinity factor (measured from only those who show a non-zero coefficient) $\geq 6\%$, above the security threshold (5%). More alarming is the fraction of the crew showing inbreeding, about 19.74% at the end of the simulation: this means that about one-fifth of the crew has a deleterious inbreeding coefficient.

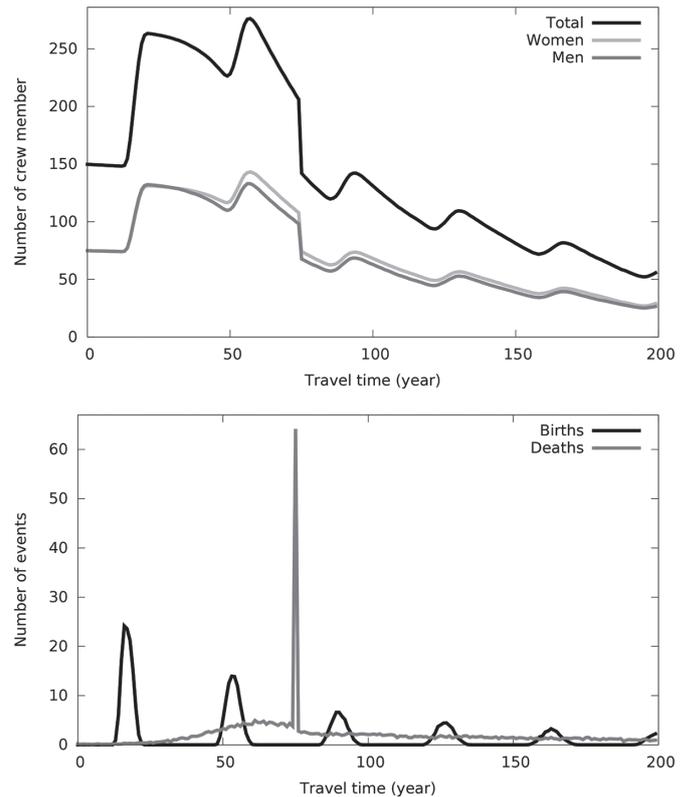

**Fig. 6** (Top) Crew evolution in terms of population number for a 200 years long trip using the parametrization from Moore [10]. (Bottom) Number of births and deaths per year. The sharp peak is due to a catastrophic sudden disaster.

In short, the numbers suggested by Moore allow a multi-generational crew to cover a 200 years long journey even if a sudden disaster impacts the space ship. The demographic sampling of the population thanks to well-defined social engineering concepts works as expected but cannot maintain a genetically healthy crew until the end of the mission. In order to help the offspring to develop without genetic disorders, the social principles must be re-assessed at each generation, according to the needs of the crew. As an example, a selective breeding program could potentially fully restore the genetic health of the population. We will develop and discuss this option in Sect. 4.

### 3.3 Results for a Smith-Like Population

The last model to be investigated is the one presented by Smith [12]. In his paper, statistics applied to large numbers and a unique MATLAB simulation are used to check whether a crew can be healthy after a 150 years long trip. Smith found that it is possible only if the founding population number lies between 14000 and 44000 crew members. In comparison to Moore studies, Smith accounted for effects of mutation, migration, selection and drift, and found that a larger initial population is necessary to compensate genetic catastrophes. To test his predictions, we thus put an initial crew of 7000 women and 7000 men in the code and increased the colony ship capacity to 50000, in order to have a similar crew/capacity ratio with the two previous studies of this paper. All other parameters are the same as discussed in Sect. 3.2, see the parametrization in Table 2, last column. Since the number of crew members to evolve in HERITAGE is much higher, the computer calculation took a longer time, about 16 hours and 15 minutes on a 4 cores Intel(R) Core(TM) i3-2100 Processor @ 3.10 GHz processor.

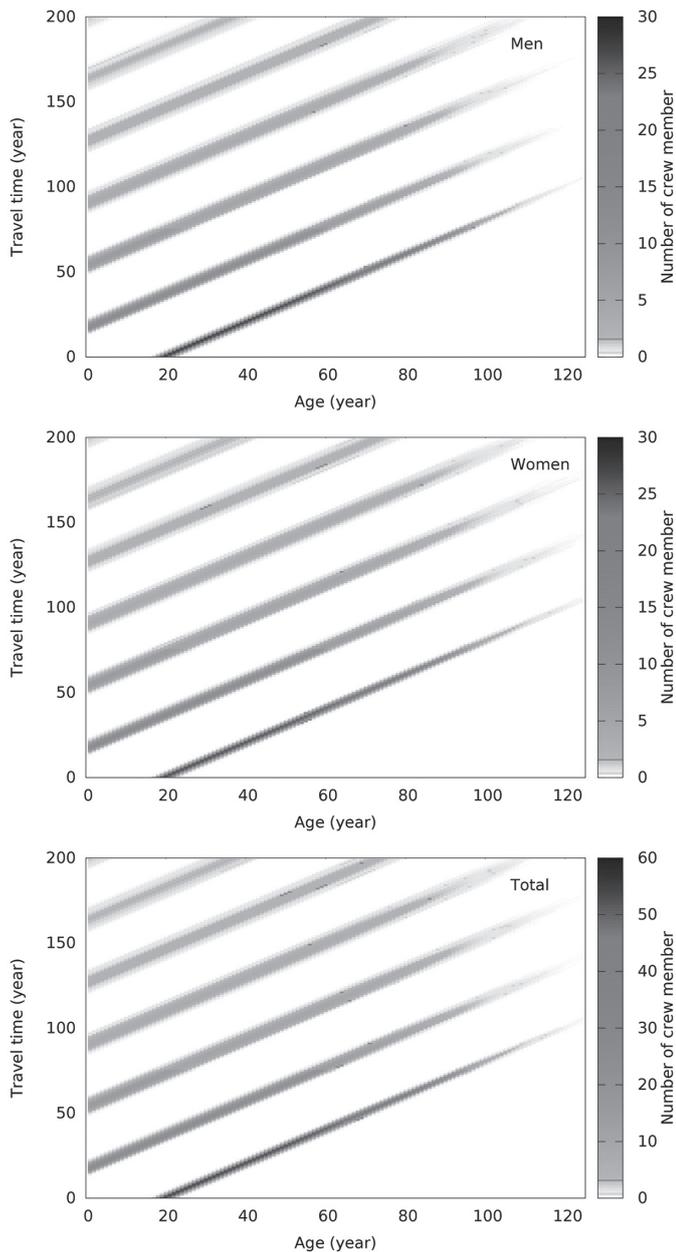

**Fig. 7** Density of crew members of a given age (x-axis) over time (y-axis) for a 200 years long trip using the parametrization from Moore [10]. The population density is colour-coded from 0 (white) to maximum (black). (Top) Men; (Middle) Women; (Bottom) Total.

We plot in Fig. 10 the evolution of this crew over 6 new generations. Except for the higher numbers, the evolution of the Moore-like population and the prediction by Smith is very similar, albeit the sharp variation due to the catastrophic event (that we selected to occur at the same time as in the two previous cases, but the reader must bear in mind that it can happen anytime over the whole 200 years). At best the ship is only at about 54.25% of its maximum capacity and, similarly to the Moore-like scenario, the population shows a steady decline to end the journey with about 6171 people, which represents ~44.08% of the initial crew. Compared to the fraction of humans reaching destination in the section above (~37.33%), the Smith-like population appears to be better resilient to a long voyage. However, the lack of revision of the initial social engineering principles throughout the journey causes the number of deaths to be unbalanced by the number of births. It is a consequence highlighted by the Monte Carlo approach, as not every woman

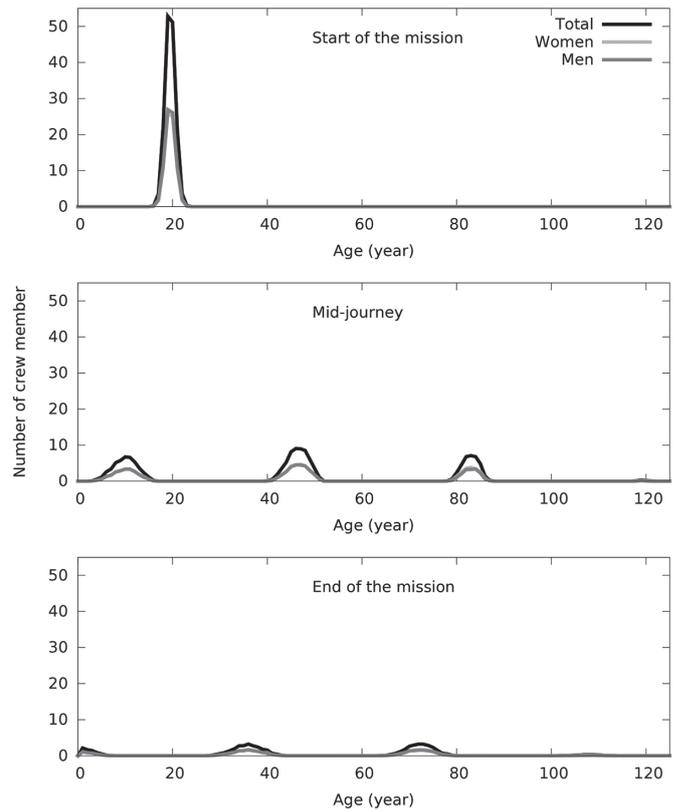

**Fig. 8** 2D slices of the density diagrams shown in Fig. 7 at three different epochs. (Top) The onset of the mission; (Middle) Half of the journey; (Bottom) End of the journey. Model: 200 years long trip using the parametrization from Moore [10].

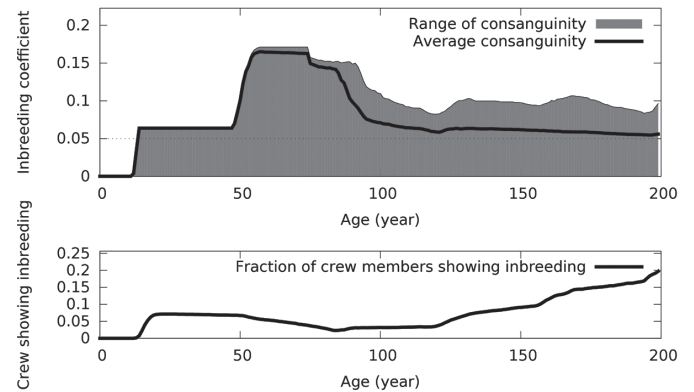

**Fig. 9** Inbreeding within the crew for a 200 years long trip using the parametrization from Moore [10]. (Top) Inbreeding coefficient as a function of time. The range of consanguinity (maximum-minimum) is show in grey and the average consanguinity factor $F$ per crew member is shown using the solid black line. The mean is measured from only those who show a non-zero coefficient. The dotted line represents the limit where deleterious effects onset. (Bottom) Fraction of crew members showing a non-zero consanguinity.

will give birth to two offspring while every male and female will eventually die.

Figures 11 and 12 are very similar to the figures shown for a Moore-like simulation. Better statistics enable us to see the fine details of the density plot of crew members. This model is thus as good as the previous one to cluster people in distinct echelons, minimizing the number of non productive people on-board and maintaining genetic variation by lengthening the generations.

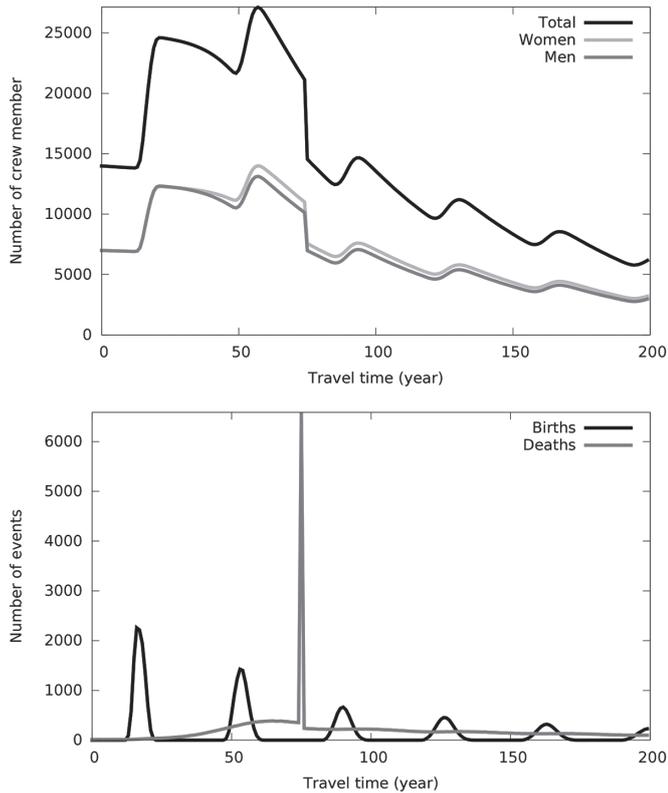

**Fig. 10** (Top) Crew evolution in terms of population number of a 200 years long trip using the parametrization from Smith [12]. (Bottom) Number of births and death per year. The sharp peak is due to a catastrophic sudden disaster.

Results are quite different regarding inbreeding. While the consanguinity coefficient rises sharply after the second generation (12.38% at maximum), the averaged inbreeding coefficient drops to ~6% at the third generation and stabilizes until the end of the journey. More importantly, the fraction of crew members showing inbreeding is very small in comparison to a Moore-like population (< 0.22%), which is negligible. The gene diversity from a much larger initial crew prevents the appearance of a large inbred subset of the crew.

The Smith-like population appears to be more efficient in mixing the genetic pool in order to ensure a safe sixth-generation to arrive on an exoplanet without severe inbreeding. The large number of initial crew members is more adapted to resist a sudden disaster and even with a severe catastrophe the viability of the mission is not compromised. However, there is a steadily decline of the population that indicates a potential risk for the long-term health of the colony. This is due to the non-adaptivity of the initial social engineering principles that cannot counterbalance the number of deaths and births.

## 4. CONCLUSIONS AND FURTHER DEVELOPMENT

In this paper, we presented HERITAGE, the first dedicated Monte Carlo code to compute the probabilistic events that will simulate the evolution of a kin-based crew aboard an interstellar ship. The code is a tool to explore if potential scenarios are viable - that is, whether a crew of a proposed size could survive without any sort of procreation regulations or artificial stocks of genetic material. As a primary test, we explored three different scenarios: 1) a population that is not under a strict birth/population control, 2) a Moore-like (small crew) population, and 3) a Smith-like (large crew) population. The first scenario fails to reach the destination due to overcrowding. Without birth control or population limits, the ship's crew capacity is exceeded after two generations. Starvation and internal conflicts will lead to the failure of this scenario. The second scenario successfully reaches destination with a reduced crew (about a third of the initial crew) but inbreeding within the crew members drives about one-fifth of the settlers to be genetically unhealthy. It is also very unlikely that a Moore-like population can survive much longer space voyages[6]. Finally, the third scenario is the only one to achieve the goal of the mission: bringing a genetically healthy crew to another distant planet, despite the fact that a small percentage (< 0.22%) of the spatial settlers have a non-zero inbreeding coefficient.

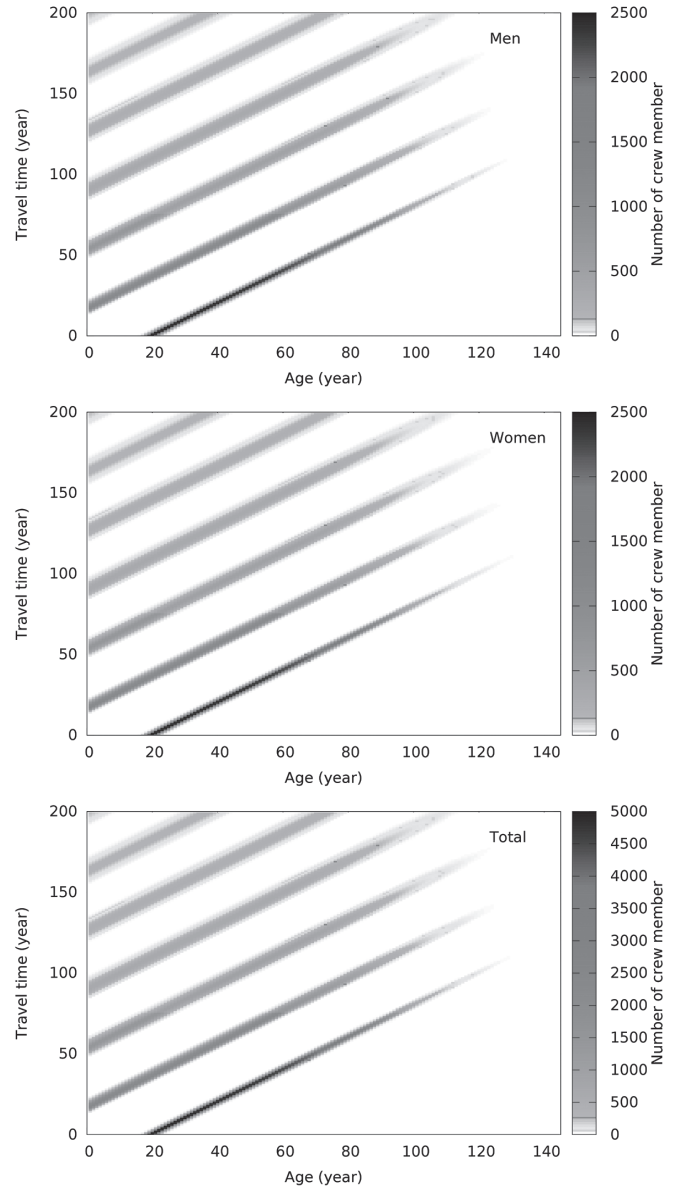

**Fig. 11** Density of crew members of a given age (x-axis) over time (y-axis) for a 200 years long trip using the parametrization from Smith [12]. The population density is colour-coded from 0 (white) to maximum (black). (Top) Men; (Middle) Women; (Bottom) Total.

The code is not programmed to maximize genetic diversity. It can control the maximum inbreeding coefficient for an offspring and limit the consanguinity of each couple, hence preventing dangerous breeding, but in this first version a maximization of the gene diversity was not explored. This will be the topic of a future improvement of the code, following the method presented in Smith [12]. Nevertheless, we have found that both a Moore and a Smith scenario are able to generate a multi-generational crew that can safely survive for 200 years. However, the first model suffers from inbreeding at a dangerous level: about 20% of its crew has a non-zero inbreeding coefficient. In order to prevent the establishment of an interstellar colony whose settlers are genetically unhealthy, either revisions must be made to the initial social engineering principles or an alternative solution is to be found in order to create a new and genetically healthy population.

---

6. We ran a 500 years long simulation and found out that, on average, a Moore-like population died towards the 470th year.

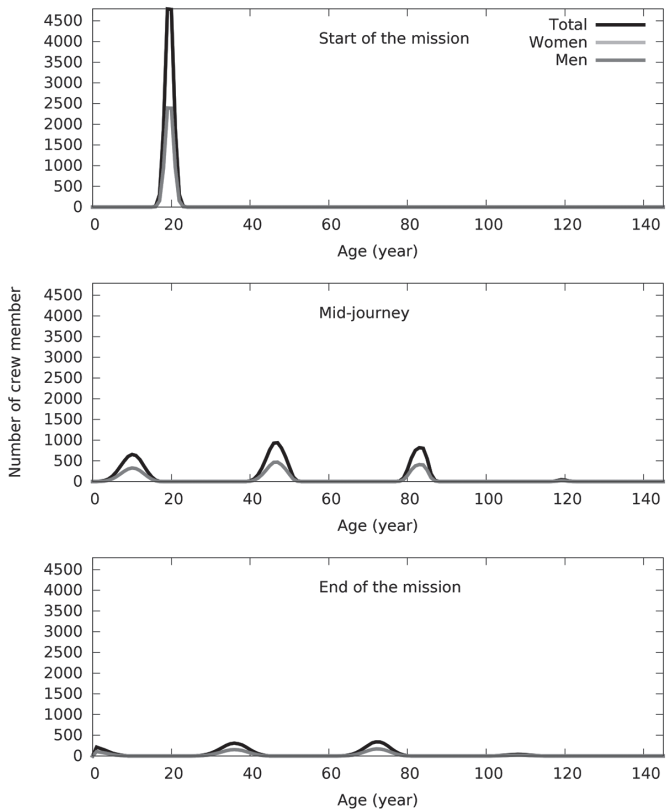

**Fig. 12** 2D slices of the density diagrams shown in Fig. 7 at three different epochs. (Top) The onset of the mission; (Middle) Half of the journey; (Bottom) At the end. Model: 200 years long trip using the parametrization from Smith [12].

As noted in the introduction, viable genetic material can be frozen and preserved for long durations (in principle indefinitely, see [42], though there could be a failure rate over time). This means that even a tiny crew could carry enough genetic material to entirely prevent inbreeding even on long timescales, and would avoid the potential difficulties of creating artificial wombs and the problems of a lack of human parents to raise the children [43]. Sociologically there's never been a population born entirely from in vitro fertilization (IVF), which also has a much greater probability of multiple births [44, 45] (though, by staggering when each crew member is fertilized, the population numbers can still be controlled). Although IVF is a complex medical procedure which may not be suitable aboard a very small ship, fertilization from stored sperm is very much simpler. However, it may not be desirable for the mission to have to rely on this method - either sociologically or technologically. HERITAGE has allowed us to investigate several proposed scenarios and evaluate whether extra generic material would be essential for the mission or if simple population controls would be sufficient.

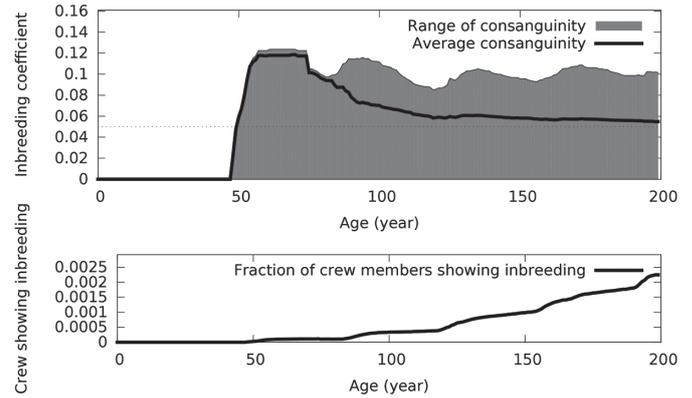

**Fig. 13** Inbreeding within the crew for a 200 years long trip using the parametrization from Smith [12]. (Top) Inbreeding coefficient as a function of time. The range of consanguinity (maximum-minimum) is show in grey and the average consanguinity factor $F$ per crew member is shown using the solid black line. The mean is measured from only those who show a non-zero coefficient. The dotted line represents the limit where deleterious effects onset. (Bottom) Fraction of crew members showing a non-zero consanguinity.

Contrary to what was predicted, all three scenarios are resistant to a sudden disaster that destroys a third of its population. The establishment of social engineering principles can easily compensate for the loss of population. Yet, it is unsure how a real crew would agree with drastic measures. In addition, the Moore-like and Smith-like scenarios have a clear decreasing trend over the trip duration, meaning that the initial social engineering principles should not be carved in stone but must evolve according to the population level at a given time. A potential consequence of this conclusion is that either a hierarchical organization or the entire ship community should work on regular revisions of the initial principles. The need for enforced social rules might drive debates and delays in defining the new social principles if a democratic vote is applied, as there would probably be a panel of non-compatible suggestions. The question of a hierarchical structure has strong implications, as it can result in a military structure with representatives of the non-democratic selection of the breeding principles. However, the social order of the spacecraft is beyond the scope of the paper and we therefore limit ourselves to mentioning this problem.

In conclusion, this paper has shown that a Moore-like or a Smith-like population are both viable prospects, especially if the starship includes a cryogenic bank of frozen sperm, eggs or embryos to ensure the creation of a colony with a new generation of genetically healthy settlers. However, strict social engineering principles are driving the population of the vessel towards extinction. There is a necessity for adaptive (i.e. to be re-assessed every generation) social engineering principles. This, together with a more realistic simulation of an interstellar travel to Proxima Centauri b, will be the subject of the following paper based on HERITAGE simulations.

**ACKNOWLEDGMENT**

The author would like to thank the anonymous referees for their constructive comments that helped to clarify this paper. I also want to thank Dr. Rhys Taylor (astrophysicist) and Dr. Camille Beluffi (particle physicist) for their careful reading of the manuscript and for their numerous comments that helped to improve this article.